\documentclass{Fernandez-Trincado}
\usepackage{graphicx}
\usepackage{xcolor}
\usepackage[colorlinks = true, citecolor = red]{hyperref}
\usepackage[]{natbib}  
\usepackage{epstopdf}

\def\BibTeX{{\rm B\kern-.05em{\sc i\kern-.025em b}\kern-.08em
    T\kern-.1667em\lower.7ex\hbox{E}\kern-.125emX}}
\bibpunct{(}{)}{;}{a}{}{,}  %%%%%%%%%%%%%  A&A bibliography style
\begin{document}

\TitreGlobal{proceeding of the SF2A 2017. 7 pages, 2 Fig}

\title{Abundance anomalies in red giants with possible extragalactic origins unveiled by APOGEE-2}

\runningtitle{Milky Way bulge}

\author{J. G. Fern\'andez-Trincado$^{1,}$}\address{Departamento de Astronom\'\i a, Casilla 160-C, Universidad de Concepci\'on, Concepci\'on, Chile, email: {\tt \url{jfernandezt@astro-udec.cl} and/or \url{jfernandezt87@gmail.com}}}\address{Institut Utinam, CNRS UMR 6213, Universit\'e Bourgogne-Franche-Comt\'e, OSU THETA Franche-Comt\'e, Observatoire de Besan\c{c}on, BP 1615, 25010 Besan\c{c}on Cedex, France, email: {\tt \url{jfernandez@obs-besancon.fr} } }

\author{D. Geisler$^{1}$}
\author{E. Moreno}\address{Instituto de Astronom\'ia, Universidad Nacional Aut\'onoma de M\'exico, Apdo. Postal 70264, M\'exico D.F., 04510, M\'exico}
\author{O. Zamora$^{4,}$}\address{Instituto de Astrof\'{\i}sica de Canarias, V\'{\i}a L\'actea S/N, E-38205 La Laguna, Tenerife, Spain}\address{Departamento de Astrof\'{\i}sica, Universidad de La Laguna (ULL), E-38206 La Laguna, Tenerife, Spain}
\author{A. C. Robin$^{2}$}
\author{S. Villanova$^{1}$}

%% Keep this line, even if the page will be settled afterwards.
\setcounter{page}{237}
%%-----------------------------------------------------------------

\maketitle

%%-----------------------------------------------------------------
%%        The abstract
%% 
%%  Warning!  within the abstract:
%%  - do not use macros. 
%%  - do not use commands like: \cite, \citet, \citep ... etc.

\begin{abstract}
By performing an orbital analysis within a Galactic model including a bar, we found that it is plausible that the newly discovered stars that show enhanced Al and N accompanied by Mg underabundances may have formed in the outer halo, or were brought in by satellites field possibly accreted a long time ago. However, another subsample of three N- and Al-rich stars with Mg-deficiency are kinematically consistent with the inner stellar halo. A speculative scenario to explain the origin of the atypical chemical composition of these stars in the inner halo is that they migrated to the inner stellar halo as unbound stars due to the mechanism of bar-induced resonant trapping.
\end{abstract}

%% Insert the keywords (to appear in the ADS indexing)
%% Keywords must be separated by a comma

\section{Introduction}

The classical picture of abundance anomalies as unique signatures of globular cluster environments have been challenged by the recent discoveries of field stars with globular cluster like abundance patterns, which become part of the general stellar population of the Milky Way. To date, only a handful of these chemically anomalous stars \citep{Ramirez2012, Majewski2012, Lind2015, Fernandez-Trincado2016b, Schiavon2017a, Recio-Blanco2017} exhibiting interesting variations in their light-element abundance patterns (e.g., C, N, O, Al, Mg, Si, and other) strikingly similar to those observed in the so-called \textit{second-generation} globular cluster stars \citep[see][for instance]{Meszaros2015, Tang2017, Schiavon2017b, Pancino2017} have been found in the Milky Way. Often, these stars have been hypothesised to be stellar tidal debris of surviving/defunct Galactic globular clusters \citep[e.g.,][]{Fernandez-Trincado2013, Kunder2014, Fernandez-Trincado2015a, Fernandez-Trincado2015b, Fernandez-Trincado2016a, Anguiano2016}. Following this line of investigation, \citet{Fernandez-Trincado2017} have recently discovered a new \textit{SG-like} stellar population across all main components (bulge, disk, and halo) of the Milky Way, with atypically low Mg abundances; i.e., similar to those seen in the \textit{second-generation} of Galactic globular cluster stars at similar metallicities. Based on the complex chemistry of these stars and their orbital properties, the authors speculate that probably most of these atypical stars may have an extragalactic origin. For example, they could be former members of dissolved extragalactic globular clusters \citep[e.g.,][]{Mucciarelli2012} and/or the result of exotic binary systems, or perhaps former members of a dwarf galaxy (with intrinsically lower Mg) polluted by a massive AGB star. 

In this work we examined the combination of proper motions from UCAC-5 \citep{Zacharias2017}, radial velocity \citep[APOGEE-2/DR14,][]{APOGEEDR142017} and spectro-photometric distances \citep[e.g.,][]{Anders2017} from the APOGEE survey, to determine the orbits of the enigmatic giant stars in a realistic Galactic potential.

\section{The Orbits}

On the basis of the absolute proper motions from UCAC-5, radial velocity and distance from the APOGEE survey, we performed a numerical integration of the orbits for five of the N-, Al-rich and Mg-poor stars studied by \citet{Fernandez-Trincado2017} in a barred Milky Way model. We employ the galactic dynamic software \textit{GravPot16}\footnote{\url{https://gravpot.utinam.cnrs.fr}} in order to carry out a comprehensive stellar orbital study. For the computations in this work, we have adopted a three dimensional potential; made up of the superposition of composite stellar populations belonging to the thin and thick disks, the interstellar medium (ISM),  the stellar halo, the dark matter halo, and a rotating bar component, that fits the structural and dynamical parameters of the Milky Way \citep[see also][]{Fernandez-Trincado2017b}. For a more detailed description of the model, we refer the reader to a forthcoming paper (Fern\'andez-Trincado et al. in preparation). For reference, the 3-dimensional solar velocity and velocity of the local standard of rest adopted by this work is: [$U_{\odot}, V_{\odot}, W_{\odot}$] = [10., 11., 7.] km s$^{-1}$, V$_{\rm LSR} =238 $ km s$^{-1}$ and the Sun is located at $R_{\odot}$ = 8.3 kpc \citep[e.g.,][]{Bland-Hawthorn2016}. For our computations, the bar pattern speed, $\Omega_{bar} = 45 $ km s$^{-1}$ kpc$^{-1}$, the total mass for the bar$=$ 1.1$\times$10$^{10}$ M$_{\odot}$, and $\phi_{bar}=20^{\circ}$ for the present-day orientation of the major axis of the Galactic bar are assumed. 

For each star, we time-integrated backwards half million orbits for 5 Gyr under variations of the proper motions, radial velocity and distance according to their estimated errors, assumed to follow a Gaussian distribution. The input parameters are listed in Table \ref{table1}. It is important to note that the uncertainties in the orbital predictions are primarily driven by the uncertainty in the proper motions and distances with a negligible contribution from the uncertainty in the radial velocity.

\begin{figure}[ht!]
 	\centering
 	\includegraphics[width=1.0\textwidth,clip]{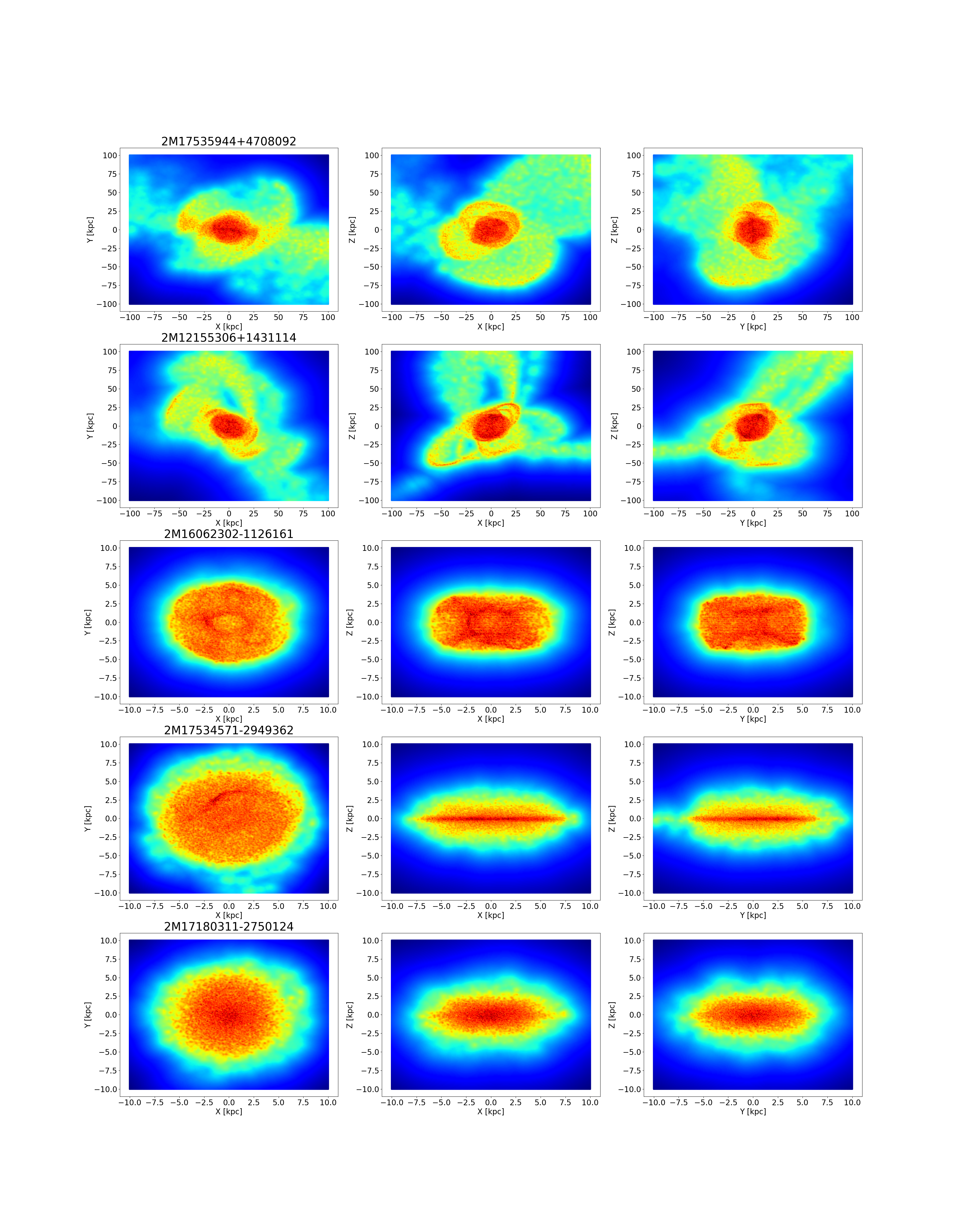}%      
 	\caption{Probability density in the equatorial Galactic plane (column 1) and side-on (column 2 and 3) of a half million simulated orbits of five chemically anomalous giant stars time-integrated backwards for 5 Gyr. Red and yellow colors correspond to larger probabilities.}
 	\label{figure1}
 \end{figure}

Figure \ref{figure1} shows the probability densities of the resulting orbits projected on the equatorial and meridional Galactic planes in the non-inertial reference frame where the bar is at rest. The red and yellow colors correspond to more probable regions of the space, which are crossed more frequently by the simulated orbits. We found that most of our stars are situated in the inner and outer halo region, which means these stars are on highly elliptical orbits (with eccentricities greater than 0.5) reaching out to a maximum distance from the Galactic plane larger than 10 kpc (outer halo) and $\sim$ 3 kpc (inner halo). The larger distances reached by 2M17535944+4708092 and 2M12155306+1431114 as well as their atypical chemical properties \citep[see][]{Fernandez-Trincado2017} suggest that these stars could have originated outside the Milky Way, brought in by satellites possibly accreted a long time ago, while the stars, 2M16062302-1126161, 2M17534571-2949362 and 2M17180311-2750124 could have migrated to the inner halo due to a mechanism known as \textit{bar-induced} resonant trapping as introduced by \citep{Moreno2015}  and became part of the general stellar population of the inner stellar halo.

  \begin{table*}[ht!]
  	\setlength{\tabcolsep}{1.7mm}  
  	\centering
  	\caption{Initial conditions of the stars analysed in this work. }
  	\label{table1}
  	\begin{tabular}{@{}ccccccccccrrcc@{}}
  		\hline
  		\hline
  		APOGEE star       & $\alpha$  & $\delta$ & distance  & $\mu_{\alpha} \times \cos (\delta)$  &  $\mu_{\delta}$  &  radial velocity    \\         
  		&        [degrees]         &        [degrees]         &        [kpc]           &  [mas yr$^{-1}$]   &   [mas yr$^{-1}$] & [km s$^{-1}$]\\
  		\hline
  		\hline    
           2M17535944+4708092 & 268.498  & 47.136  & 15.35$\pm$1.7 &  -4.1$\pm$2.2 & -1.1$\pm$2.2  & -266.02$\pm$0.02  \\
           2M12155306+1431114 & 183.971  &    14.519     & 14.27$\pm$1.54    & 0.3$\pm$3.1       &   -2.6$\pm$3.1    & 100.08$\pm$0.01   \\
           2M16062302-1126161 & 241.596   & -11.438     &   3.58$\pm$0.32    & -6.1$\pm$1.0      &  -8.5$\pm$1.0    &  -105.90$\pm$0.01  \\
           2M17534571-2949362 & 268.440   & -29.827     &   3.38$\pm$0.75    &  -6.3$\pm$2.5     &  -8.4$\pm$2.5     & -140.68$\pm$0.02   \\
           2M17180311-2750124 & 259.513   & -27.837     &   4.75$\pm$1.18    & -10.0$\pm$2.2    &  -9.0$\pm$2.2     & -113.71$\pm$0.03   \\                      
  		\hline 
  		\hline
  	\end{tabular}
  \end{table*}

\section{Concluding Remarks}

Orbital modelling shows that two of our chemically anomalous stars (2M17535944+4708092, 2M12155306+1431114) have a large motion out of the plane of the Milky Way, with $Z_{max} > 10 $ kpc. The eccentricity of the orbit and this relatively large out-of-plane motion suggest either an extra-galactic origin, or star formation at large Galactic distances. While three stars in our sample (2M16062302-1126161, 2M17534571-2949362, 2M17180311-2750124) are kinematically consistent with the inner stellar halo which is thought to have higher eccentricities, we argue that this subsample may come from stars born in the outer halo and kinematically heated into the inner Galactic halo due to perturbations by resonances with the Galactic bar \citep[e.g.,][]{Moreno2015}. The orbital projections presented here confirm the assumptions of our previous work \citep{Fernandez-Trincado2017} that most of these atypical stars belong to the outer halo, possibly brought in by extragalactic stellar systems.

Lastly, the future \textit{Gaia} data releases should improve the precision to which the star's orbit can be calculated by providing accurate and precise proper motions and parallax.

% Optional acknowledgements
% -------------------------
\begin{acknowledgements}
	
J.G.F-T and D.G. gratefully acknowledges partial financial support from the SF2A in order to attend the SF2A-2017 meeting held in Paris in July 2017, and the Chilean BASAL Centro de Excelencia en Astrof\'isica y Tecnolog\'ias Afines (CATA) grant PFB- 06/2007. E.M acknowledge support from UNAM/PAPIIT grant IN105916. O.Z. acknowledge support provided by the Spanish Ministry of Economy and Competitiveness (MINECO) under grant AYA-2014-58082-P. We also acknowledge the support of the UTINAM Institute of the Universit\'e de Franche-Comt\'e, R\'egion de Franche-Comt\'e and Institut des Sciences de l'Univers (INSU) for providing HPC resources on the Cluster Supercomputer M\'esocentre de calcul de Franche-Comt\'e. 

 Funding for the \textit{GravPot16} software has been provided by the Centre national d'\'etudes spatiale (CNES) through grant 0101973 and UTINAM Institute of the Universit\'e de Franche-Comte, supported by the Region de Franche-Comte and Institut des Sciences de l'Univers (INSU). 
\end{acknowledgements}

\end{document}